\documentclass[twocolumn,tighten,times]{aastex62}

\hypersetup{linkcolor=black,citecolor=black,filecolor=cyan,urlcolor=magenta}

\newcommand{\um}{$\mu$m}

\accepted{ApJL}
\shorttitle{The highly polarized dusty emission core of Cygnus A}
\shortauthors{Lopez-Rodriguez et al.}
\begin{document}

\title{The highly polarized dusty emission core of Cygnus A}

\correspondingauthor{Lopez-Rodriguez, E.}
\email{elopezrodriguez@sofia.usra.edu}

\author{Enrique Lopez-Rodriguez}
\affil{SOFIA Science Center, NASA Ames Research Center, Moffett Field, CA 94035, USA}
\affil{National Astronomical Observatory of Japan, National Institutes of Natural Sciences (NINS), 2-21-1 Osawa, Mitaka, Tokyo 181-8588, Japan}

\author{Robert Antonucci}
\affil{Department of Physics, University of California in Santa Barbara, Broida Hall, Santa Barbara, CA 93109, USA}

\author{Ranga-Ram Chary}
\affil{IPAC/Caltech, MS314-6, Pasadena, CA 91125, USA}

\author{Makoto Kishimoto}
\affil{Department of Physics, Kyoto Sangyo University, Motoyama, Kamigamo, Kita-ku, Kyoto, 603-8555, Japan}



\begin{abstract}

We report the detection of linearly polarized emission at 53 and 89 \um~from the radio-loud active galactic nucleus (AGN) Cygnus A using  HAWC+ onboard SOFIA. We measure a highly polarized core of $11\pm3$\% and $9\pm2$\% with a position angle (P.A.) of polarization of $43\pm8^{\circ}$ and $39\pm7^{\circ}$ at 53 and 89 \um, respectively.  We find (1) a synchrotron dominated core with a flat spectrum ($+0.21\pm0.05$) and a turn-over at $543\pm120$ \um, which implies synchrotron emission is insignificant in the infrared (IR), and (2) a $2-500$ \um\ bump peaking at $\sim40$ \um~described by a blackbody component with color temperature of $107\pm9$ K. The polarized SED has the same shape as the IR bump of the total flux SED.  We observe a change in the P.A. of polarization of $\sim20^{\circ}$ from 2 to 89 \um, which suggests a change of polarization mechanisms. The ultraviolet, optical and near-IR polarization has been convincingly attributed to scattering by polar dust, consistent with the usual torus scenario, though this scattered component can only be  directly observed from the core in the near-IR.  By contrast, the gradual rotation by $\sim20^{\circ}$  towards the far-IR, and the near-perfect match between the total and polarized IR bumps, indicate that dust emission from aligned dust grains becomes dominant at $10-100$ \um, with a large polarization of 10\% at a nearly constant P.A. This result suggests that a coherent dusty and magnetic field structure dominates the $10-100$ \um\ emission around the AGN.

\end{abstract}



%%%%%%%%%%%%%%%%%%%
%%%%% INTRODUCTION %%%%%
%%%%%%%%%%%%%%%%%%%

\section{Introduction} \label{sec:int}

Cygnus A \citep[z =  0.0562,][H$_{o}$ = 73 km s$^{-1}$ Mpc$^{-1}$; 1\arcsec~$\sim$~1 kpc]{1994AJ....108..414S}  is the most studied Faranoff-Riley II (FRII) radio galaxy. This galaxy shows a complex central few kpc (patchy dust lane, dusty ionization cone, and jets), with a heavily extinguished core.  Moderate (few arcsec; few kpc) and sub-arcsecond (few hundred pc) angular resolution observations of the nucleus of Cygnus A have provided unclear dominant physical components. \citet{2012ApJ...747...46P} suggested that star forming regions in the far-infrared (FIR), a thermal component at near- and mid-IR wavelengths, and a synchrotron component at radio wavelengths are the main physical components at scales of few kpc. However, \citet{Koljonen:2015aa} suggest that the core is  dominated by a synchrotron component with a turn-over in the IR wavelength range at scales of a few hundred pc.
 
Polarimetric techniques provide an alternative approach to study the core and surrounding environments of AGN. Optical \citep{1990MNRAS.246..163T,1997ApJ...482L..37O} and ultraviolet \citep[UV;][]{Hurt:1999aa} polarimetric observations show that the dominant polarization mechanism arises from scattering in the extended cones, while  the central few kpc is obscured at these wavelengths. This result is in agreement with the unified model of active galactic nuclei \citep{Antonucci:1993aa,Urry:1995aa}. However, due to the kpc-scale dust lane in Cygnus A, and other radio sources, optical and UV observations do not provide significant information about the core due to extinction \citep[i.e.][]{1990ApJ...363L..17A}. IR and (sub-)mm observations are required to study the core of radio galaxies.

\citet{Ramirez:2014aa} performed 2.05 \um~imaging polarimetric observations of 13 FRII radio sources using NICMOS/\textit{HST} and found a generally high intrinsic polarized ($6-60$\%) core, and a moderately strong tendency for the position angle (P.A.) of polarization to be perpendicular to the jet direction. These authors suggested the dominant polarization mechanism at 2.05 \um~arises from dichroic extinction in the core of these radio sources, although other polarization mechanisms could also be present. For Cygnus A, sub-arcsecond angular resolution 2.0 \um~observations using NICMOS/\textit{HST} by \citet{Tadhunter:2000aa}  measured a highly polarized ($\sim10$\% observed, $\sim28$\% starlight-corrected) core with a P.A. of polarization ($201\pm3^{\circ}$) nearly perpendicular to the radio jet \citep[P.A.$_{jet} \sim 284^{\circ}$,][]{Sorathia:1996aa}. These results indicate that the 2 \um\ polarization arises from scattering within the central 375 pc. 

Further sub-arcsecond resolution 10~\um~observations using CanariCam on the 10.4-m Gran Telescopio CANARIAS by \citet{Lopez-Rodriguez:2014aa} also found a highly polarized ($\sim10$ \%) core within the central 0.38\arcsec (380 pc) with a slight shift in the P.A. of polarization of $\sim10^{\circ}$ with respect to the 2 \um~observations. The change in P.A. of polarization from 2 to 10 \um~indicates a different polarization mechanism may dominate at longer wavelengths. These authors suggested that the 10~\um~polarization arises from a self-absorbed synchrotron component with a turn-over at $\sim34$ \um~attributed to the pc-scale jet close to the core. The self-absorbed synchrotron component with a break in the IR wavelength range \citep{2012ApJ...747...46P,2014ApJ...788....6M,Koljonen:2015aa,Lopez-Rodriguez:2014aa} implies that the emission arises from a single zone with highly coherent magnetic field in the jet, and with a sharp stop in production of electrons above the break frequency. Empirically, \cite{1990ApJ...353..416A} found that lobe-dominated radio galaxies cores are dominated by self-absorbed synchrotron emission with a break at mm wavelengths. This result shows a jet dominated core at long wavelengths, while a thermal mechanism dominates the core in the IR. Thus, SED modeling of Cygnus A is in contradiction with the empirical findings in lobe-dominated radio galaxies. Total and polarization flux observations in the far-IR (FIR) and mm are required to disentangle the thermal and non-thermal components in the core of radio-loud galaxies.

We present FIR imaging polarimetric observations of the core of Cygnus A, and the total and polarized nuclear SED from 2 \um~to sub-mm wavelengths. We will show that such a combination allows us to disentangle the thermal and non-thermal components in the core of Cygnus A. 

%Section \ref{sec:obs} presents the newly obtained FIR polarimetric observations and results, Section \ref{sec:sed} compiles the nuclear SED and discussed it.

%%%%%%%%%%%%%%%%%%%%%%%%%%%%%%%%
%%%%% OBSERVATIONS AND DATA REDUCTION %%%%%
%%%%%%%%%%%%%%%%%%%%%%%%%%%%%%%%

\section{The Far-Infrared polarimetric Observations} \label{sec:obs}

%%%%%%%%%%%%%
%%%% FIGURE 1 %%%%
%%%%%%%%%%%%%%
%% The "ht!" tells LaTeX to put the figure "here" first, at the "top" next
%% and to override the normal way of calculating a float position
\begin{figure}[h!]
\includegraphics[angle=0,scale=0.4]{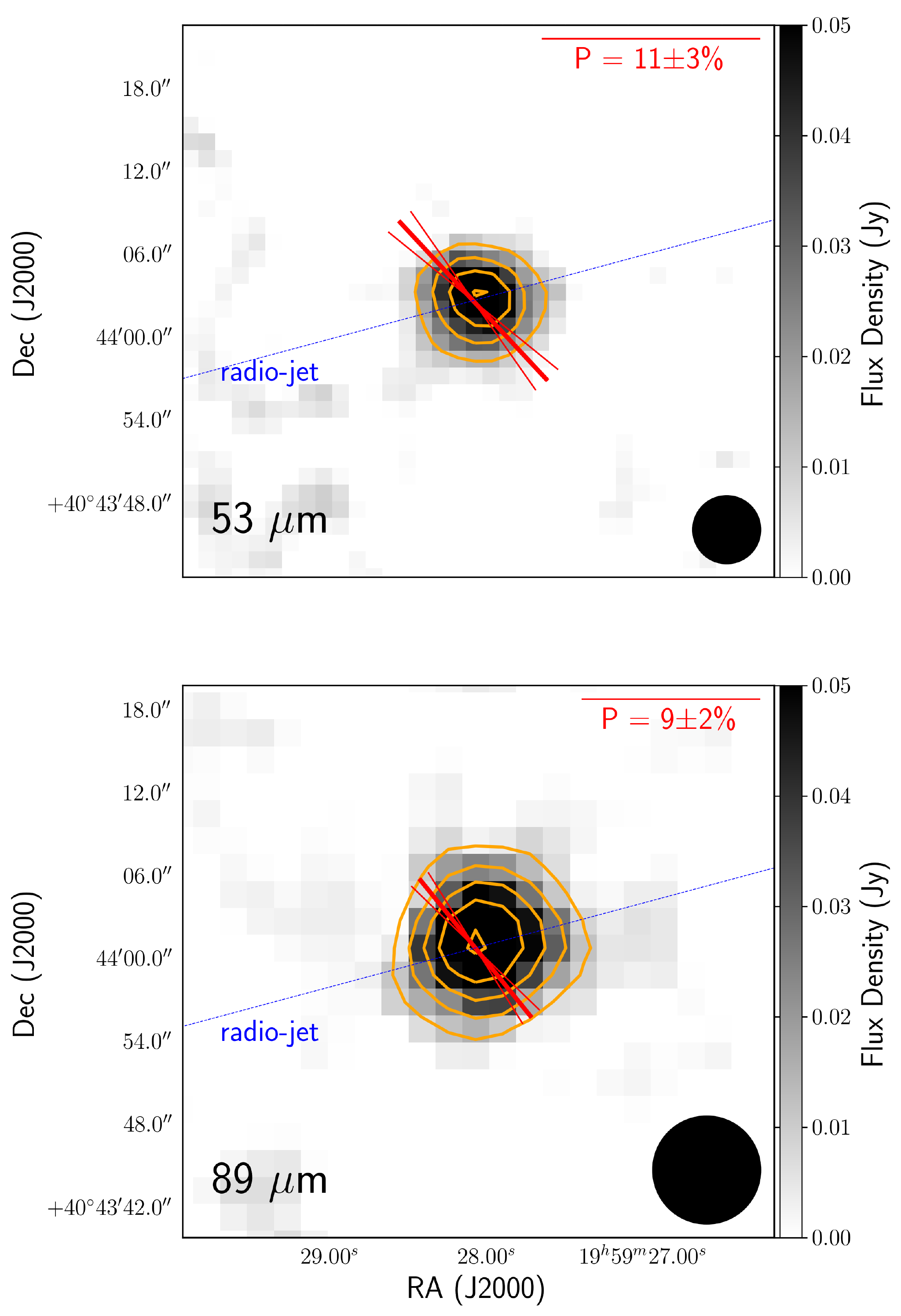}
\caption{Imaging-polarimetric observations at 53 (top) and 89 \um~(bottom). Flux (grey scale) and polarization (red vector) with the uncertainty (thin red solid vector) in the P.A. of polarization within the beam size (black circles) are shown. Orange contours start at 3$\sigma$ and increases in steps of 3$\sigma$. Radio-jet axis (blue dashed line) with a P.A.$_{jet} \sim 284^{\circ}$ \citep{Sorathia:1996aa} is shown.
\label{fig:fig1}}
\epsscale{2.}
\end{figure}
%%%%%%%%%%%%%%

Cygnus A was observed (PI: Lopez-Rodriguez, ID: 05\_0071) on 20171025/27 using the High-resolution Airborne Wideband Camera-plus (HAWC+) \citep[][Harper et al., submitted]{Vaillancourt:2007aa} on the 2.5-m Stratospheric Observatory For Infrared Astronomy (SOFIA) telescope. We made observations using the chop-nod polarimetric mode at 53 \um~($\lambda_{C} = 53$ \um, $\Delta \lambda / \lambda_{c} = 0.17$ bandwidth) and 89 \um~($\lambda_{C} = 88.7$ \um, $\Delta \lambda / \lambda_{c} = 0.19$ bandwidth). HAWC+ polarimetric observations simultaneously measure two orthogonal components of linear polarization arranged in two pair of arrays of $32 \times 40$ pixels each, with angular resolution of 2.55\arcsec\ and 4.02\arcsec\ per pixel at 53 and 89 \um~respectively. The beam size has been estimated to be 4.85\arcsec\ and 7.80\arcsec\ at 53 and 89 \um\, respectively.

In both bands, we performed observations in a four dither pattern with an offset of three pixels, where four halfwave plate P.A. were taken in each dither position. We used nod times of 35s and 40s at 53 and 89~\um, respectively, with a chop frequency of 10.2 Hz, chop-amplitude of 90\arcsec, and chop-angle of 90$^{\circ}$ with respect to the short axis of the array. Based on the morphology of the source in the \textit{Herschel} observations (Section \ref{subsec:TFSED}), the chop-throw and chop-angle configuration does not result in any significant flux contribution from the radio sources and/or diffuse emission from the background. At 53 \um, six and three dither sets were taken the first and the second night of observations, respectively. Several dither sets were acquired with chop-throws of 180\arcsec, providing a slightly elongated image due to misalignments by the chop-nod of SOFIA. To minimize potential polarization contamination due to misaligned images, we removed these datasets from the analysis. At 89 \um, five dither sets were taken the second night of observations. Final observations provide a total exposure time of 1399s and 1483s at 53 and 89 \um, respectively.  Data were reduced using the \textsc{hawc\_dpr pipeline} v1.3.0beta1 and custom \textsc{python} routines. The pipeline follows the procedure as described by Berthoud et al. (in preparation). Raw data were demodulated, chop-nod and background subtracted, flux calibrated and Stokes parameters estimated with their uncertainties. In order to account for systematic or random uncertainties, we performed a reduced-$\chi^{2}$ analysis on the polarimetric data. We estimated an excess noise factor of $\sqrt{\chi^{2}/\chi^{2}_{t}} \sim 1.16$, where $\chi^{2}_{t}$ is the theoretical reduced-$\chi^{2}$ for a given set of observations. Finally, we inflated the uncertainties of the Stokes parameters such that the final excess noise factor $\sqrt{\chi^{2}/\chi^{2}_{t}} = 1$.

Figure \ref{fig:fig1} shows the imaging polarimetric observations at 53 and 89 \um. In both bands, we found a point-like source with a highly polarized ($\sim10$\%) core with a P.A. of polarization of $\sim40^{\circ}$ (Table \ref{tab:table1}). To increase the signal-to-noise ratio of the polarimetric measurements, we measure the nuclear polarization within the beam size at each band. These measurements provide a single statistically significant polarization vector at the $\sim4\sigma$ level in the degree of polarization.

%%%%%%%%%%%%%
%%%%% SED %%%%%
%%%%%%%%%%%%%

\section{Nuclear SED} \label{sec:sed}

%%%%%%%%%%%%%
%%%% FIGURE 2 %%%%
%%%%%%%%%%%%%%

%% The "ht!" tells LaTeX to put the figure "here" first, at the "top" next
%% and to override the normal way of calculating a float position
\begin{figure*}[ht!]
\includegraphics[angle=0,scale=0.52]{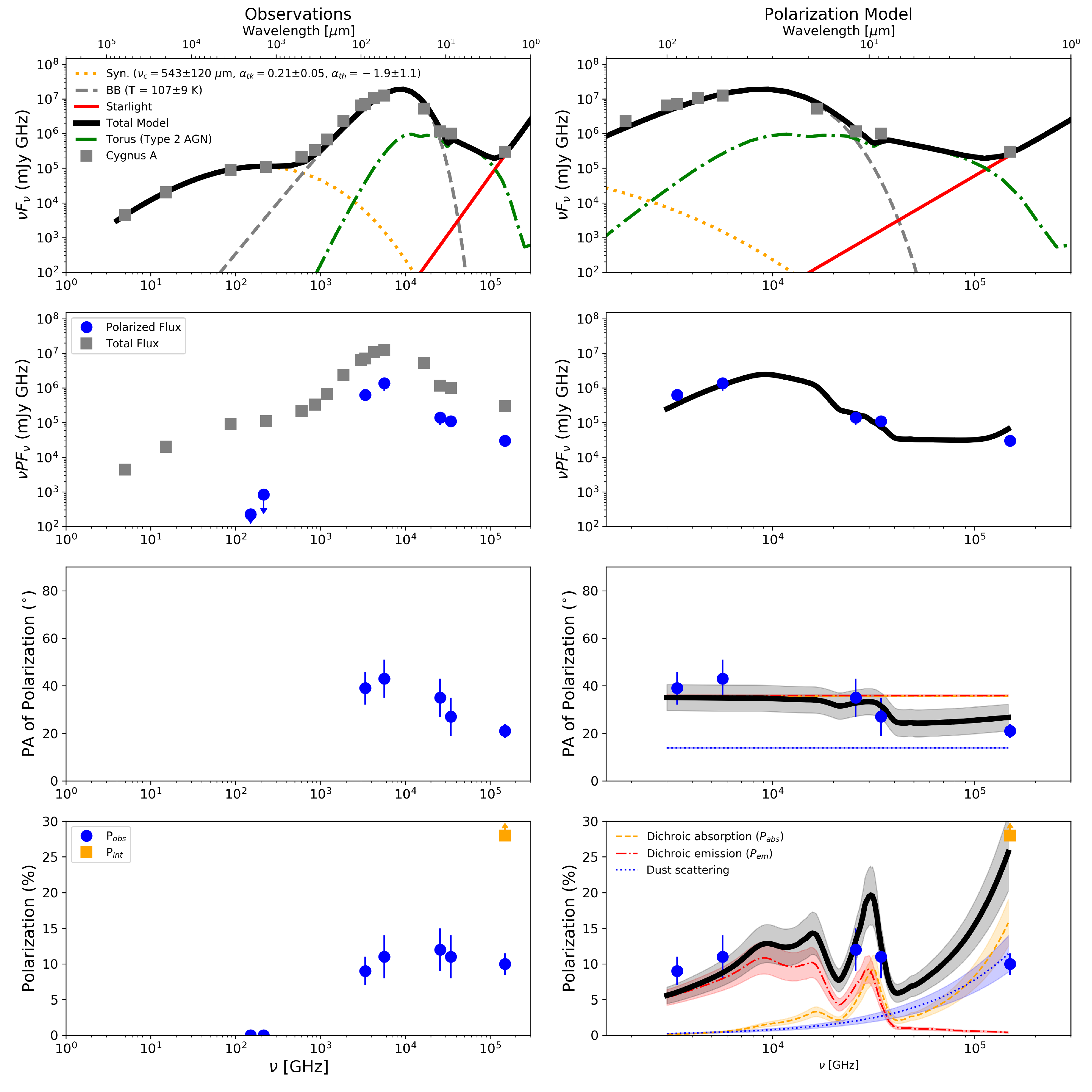}
\caption{Observations (left column) and polarization model (right column) of the nucleus of Cygnus A showing the total (first row) and polarized (second row) fluxes, P.A. (third row) and degree (fourth row) of polarization. The intrinsic (starlight corrected) nuclear polarization (orange square) at 2.2 \um~by \cite{Tadhunter:2000aa} is shown. SED model: a self-absorbed synchrotron component (orange dotted line), a blackbody component with color temperature of $107\pm9$ K (grey dashed line), a clumpy torus model of a Type 2 AGN  (green dashed line), and a starlight component (red solid line) describe the nuclear SED of Cygnus A. Polarized SED model: the combination of dichroic absorption (orange) and emission (red), and dust scattering (blue) explain the degree and P.A. of polarization. The total polarized flux model (black line) is estimated as the multiplication of the total flux model by the total polarization model. 
\label{fig:fig2}}
\epsscale{2.}
\end{figure*}

%%%%%%%%%%%%%%%%%%%%%%%%%

We compile the total and polarized nuclear SED of Cygnus A with the best spatial resolution available from 2 \um~to sub-mm wavelengths. Figure \ref{fig:fig2}  (Table \ref{tab:table1} in Appendix \ref{app:sup}) summarizes the data. For all observations, the angular resolution is below the separation of $\sim45$\arcsec\ between the core and radio lobes, which ensures that the core is isolated from the radio lobes.

\subsection{Total flux SED}\label{subsec:TFSED}

We took \textit{Herschel} observations of Cygnus A  from the \textit{Herschel} Archive\footnote{\textit{Herschel} Archive can be found at \url{http://archives.esac.esa.int/hsa/whsa/}}. To our knowledge, these data have not been published before, so we present the observations  here. At all wavelengths, the core and radio lobes are resolved, which allows us to perform photometric measurement of the core without any contamination from the radio lobes. We found that the flux density of the core decreases with increasing wavelength, while the flux density of the radio lobes increases (Fig. \ref{fig:fig3}). We did not find any contribution from the lobes in our 53 and 89 \um~HAWC+ observations, which indicates that the nuclear photometric and polarimetric measurements are dominated by a compact unresolved component.

Observationally, we find that the nuclear SED of Cygnus A shows a turn-over emission in the mm wavelength regime, and an IR bump peaking at $\sim$40 \um. We fitted a model to the  SED (Fig. \ref{fig:fig2}-top-left). We find that a self-absorbed synchrotron component with a flat-spectrum ($F_{\nu} \propto \nu^{\alpha_{tk}}$, with $\alpha_{tk} = +0.21\pm0.05$) in the optically thick region with a turn-over at $\nu_{c} = 552\pm100$ GHz ($\lambda_{c} = 543\pm120$ \um) best explains the mm range. The index of the optically thin region ($\alpha_{th} = -1.9\pm1.1$)  cannot be well constrained due to the  contribution of the IR bump. The synchrotron emission is insignificant at IR wavelengths in the core of Cygnus A. 

The IR bump is best described by a single blackbody component from 10 to 500 \um. Then, a torus component is required to account for the 10 \um\ silicate feature, and starlight is required to account for the emission at 2 \um. We use the \textsc{clumpy} torus template for the archetypical Type 2 AGN, NGC 1068, inferred using subarcsecond angular resolution observations from IR to sub-mm wavelengths \citep{2018ApJ...859...99L}. For the starlight component, we use the Elliptical galaxy template from the Spitzer Wide-area InfraRed Extragalactic survey (SWIRE) template library\footnote{The SWIRE template library can be found at \url{http://www.iasf-milano.inaf.it/~polletta/templates/swire_templates.html}} \citep{2007ApJ...663...81P}. The blackbody component has a characteristic color temperature of $107\pm9$ K  assuming a dust emissivity of $\nu^{1.6}$. Our color temperature is compatible with the estimated dust temperature of $150\pm10$ K within the central 2 kpc using N-band observations using OSCIR/Keck II \citep{2002ApJ...566..675R}. \cite{1980ApJ...235...18K} suggested, for Cygnus A, a non-thermal origin up to sub-mm wavelengths with a thermal component as the dominant components at IR wavelengths, which agrees with our well-sampled nuclear SED. These results agree with the spectra of lobe-dominant quasars having a non-thermal component with a turn-over in the sub-mm wavelength range and an IR excess \citep{1990ApJ...353..416A}. 

Our SED fitting shows that a single blackbody component does not account for both the 10 \um\ window and 2 \um\ measurements. We explain the emission in the 10 \um\ window as the mixture of contributions from the dust emission of the blackbody component and the pc-scale \textsc{clumpy} torus. \citet{2014ApJ...788....6M} found that the 10 \um\ silicate absorption feature observed using COMICS/Subaru can be explained by a blackbody component with a temperature of $217\pm3$ K obscured by cold dust  with a characteristic temperature of $<50$ K, or a \textsc{clumpy}. We explain the emission at 2 \um\ as dominated by starlight/scattered light (Section \ref{sec:NucPol}).

Using the estimated blackbody temperature of $107\pm9$ K, and the bolometric luminosity of $1 \times 10^{45}$ erg s$^{-1}$ \citep{2002ApJ...566..675R}, we estimated an extension of the dust emission to be within a radius of $17\pm3$ pc ($\sim$17 mas). \cite{2000ApJ...535..626I} suggested an obscuring dusty structure with inner radius of $2.25$ pc  and outer radius of a few hundred pc based on IR SED fitting. We estimate the dust mass at the peak frequency from a modified blackbody using, in practical units, $M_{d} = 1.6 \times 10^{-6} \nu_{1000GHz}^{-(2+\beta)} S_{\nu,Jy} T_{d,K}^{-1} D_{pc}$ M$_{\odot}$ \citep{2016A&A...587A..73B}, where $\nu_{1000GHz}$ is the frequency normalized to 1000 GHz, $\beta$ is the index of the dust opacity, $T_{d,K}$ is the dust temperature of the blackbody component in units of K, and $D_{pc}$ is the distance to the source in units of pc. We took $\nu_{1000GHz} = 2.99$, $\beta = 1.6$,  $T_{d,K} = 107\pm9$ K, and  $D_{pc} = 2.37 \times 10^{8}$  Mpc, and estimate a dust mass of $M_{d} = 3.3 \pm 0.6 \times 10^{7}$ M$_{\odot}$. \citet{2002ApJ...564..176Y} estimated a column density of gas to be $N_{H} = 2 \times 10^{23}$ cm$^{-2}$ using X-ray \textit{Chandra} observations. We took this column density, and we estimate a gas mass of $3 \times 10^{8}$ M$_{\odot}$ within a 17 pc radius \citep[][eq. 2]{2000MNRAS.318.1232W}. We finally estimate the gas-to-dust ratio to be 0.1. Using a sample of 48 galaxies in the SINGS sample, \citet{2007ApJ...663..866D} found a dust-to-gas ratio in the range of 0.002-0.005. \cite{2012MNRAS.422.2291P} estimated a dependence of the dust-to-gas ratio as a function of the distance to the central source of Centaurus A with an average of 0.001 in an area of $10 \times 2$ arcmin$^{2}$. Typical Galactic values of the gas-to-dust ratio varies from 0.006-0.008 \citep[e.g.][]{2001ApJ...554..778L,2004ApJS..152..211Z}.

\subsection{Nuclear polarization}\label{sec:NucPol}

\cite{Ritacco2017} performed 12\arcsec\ and 18.2\arcsec~resolution polarimetric observations at 1.15 and 2.05 mm of Cygnus A. They found an unpolarized core, with highly polarized, $\sim 10$\%, radio lobes with perpendicular P.A. of polarization to each other. Using their observations, we obtain an 3-$\sigma$ upper-limit of the polarized flux of the core to be $<3.9$ and $<1.5$ mJy at 1.15 and 2.05 mm, respectively. We measured the polarization of Cygnus A to be 0.4\% to 3.2\% with a fairly constant P.A. of polarization of $\sim165^{\circ}$ from 30 to 217 GHz using \textit{Planck} observations \citep{2016A&A...594A..26P}, in a 5\arcmin~aperture at 545 GHz and 30\arcmin~aperture at 30 GHz. The low angular resolution from \textit{Planck} measurements are highly affected by the polarized radio lobes, and they are not used in this work.

Observationally, we find the polarized SED (Fig. \ref{fig:fig2}-top-right) has the same shape as the IR bump of the total flux SED. Despite the angular resolution observations from 2.0 to 89 \um, the degree of polarization remains nearly constant ($\sim10$\%) within the uncertainties, although a tentative decrease in the degree of polarization from 10 to 89 \um~is observed. We measure a statistically significant change of $\sim20^{\circ}$ in the P.A. of polarization from 2 to 89 \um.  The change in the P.A. of polarization from 2 to 89 \um, and the tentative change of the degree polarization in the 10 to 89 \um~indicate a change of polarization mechanisms between 2 to 89 \um.

We investigate the several polarization mechanisms arising from the nucleus of Cygnus A from 1 to 100 \um. Specifically, we follow similar techniques as presented by \citet{Aitken:2004aa} and \citet{Lopez-Rodriguez:2016ab}, which have been proven \citep[e.g.][]{Smith:2000aa} to be useful to disentangle the several competing polarization mechanisms in the MIR. We here extend this polarization technique to the wavelength range from 1 to 100 \um. The known polarization profiles of the absorptive and emissive polarization components, as well as the wavelength-dependent cross-section dust scattering within the $1-100$ um, are linearly combined to obtain the best fit to the normalized Stokes qu. Specific details are shown in Section \ref{app:model}. As shown in Section \ref{subsec:TFSED}, synchrotron emission is insignificant in the IR, therefore this component is not used in the polarization model. Figure \ref{fig:fig2}-right shows the final polarization model for the total and polarized flux, and for the degree and P.A. of polarization. Figure \ref{fig:fig4} shows polarization profiles, and best fit to the normalized Stokes qu. 

Our best fit shows that the combination of dichroism and dust scattering best explain the wavelength-dependent degree and P.A. of polarization in the $1-100$ \um~wavelength range. Figure \ref{fig:fig4} shows the fractional contribution of each polarization mechanism per wavelength. We find that the $20-100$ \um\ polarized SED is explained by dichroic emission of aligned dust grains in the unresolved core of Cygnus A. Polarization arising by dichroic emission is the dominant polarization mechanism in the $20-100$ \um~wavelength range. In the $8-12$ \um~wavelength range, dichroic emission and dichroic absorption compete. This behavior has been found to be typical in astrophysical objects, i.e. star forming regions, young stellar objects, etc. \citep{Smith:2000aa}. At 2 \um, dust scattering polarization is extinguished by the dichroic absorption component, both having statistically significant different P.A. of polarization of $36\pm6^{\circ}$ and $14\pm5^{\circ}$, respectively. \citet{Ramirez:2014aa} found, in a sample of 13 FRII galaxies, that dichroic absorption as well as scattering can be the dominant polarization mechanisms at 2 \um, but the single wavelength polarization observation made it difficult to disentangle them. The polarization at 2 \um~is interpreted as a combination of several mechanisms, i.e. dust scattering and dichroic absorption. We note that our estimated P.A. of polarization of $14\pm5^{\circ}$ at 2 \um~arising from dust scattering is perpendicular to the radio jet axis of P.A.$_{jet} \sim 284^{\circ}$ \citep{Sorathia:1996aa}.

%%%%%%%%%%%%%%%%%%%%
%%%%% FINAL REMARKS %%%%%
%%%%%%%%%%%%%%%%%%%

\section{Final Remarks} \label{sec:dis}

As mentioned in the introduction, the UV/optical polarization \citep{1990MNRAS.246..163T,1997ApJ...482L..37O,Hurt:1999aa,Tadhunter:2000aa,2003MNRAS.345L..13V} is dominated by dust scattering arising from the few-kpc scale bicone. No information about the core is obtained due to the obscured core at these wavelengths. The fact that the P.A. of polarization remains perpendicular to the jet direction from UV to 2 \um\ \citep{1998MNRAS.297..936P,Tadhunter:2000aa,Ramirez:2014aa} and based on results from our polarization model, we suggest that the 2 \um\ polarization arises from dust scattering from an unresolved dusty scattering region within the 350 pc central region. The dichroic absorption component of our model indicates the need of extinction correction to obtain the intrinsic polarization at 2 \um. The change in the P.A. of polarization of $\sim$20$^{\circ}$ from 2 to 89 \um\ suggests a change of polarization mechanisms.

The nearly constant P.A. of polarization in the FIR, the near-perfect match between the total and polarized SED at wavelengths $>2$ \um, the estimated high dust-to-gas ratio at 17 pc using the blackbody component with color temperatures of $107\pm9$ K, and our polarization model, indicate that dichroic dust emission becomes dominant at FIR wavelengths. It also suggests a fairly coherent and well-defined, albeit unresolved, structure for this emitting dust. The MIR polarization seems to be the wavelength range where the transition of polarization mechanisms take place. Specifically, the MIR polarization arises from a contribution of absorptive and emissive polarization from aligned dust grain surrounding the AGN. We note that previous analysis \citep{Lopez-Rodriguez:2014aa} have suggested that the dominant MIR polarization arises from polarized synchrotron emission from the pc-scale jet of Cygnus A. Although their suggested power law with a cut-off at $\sim34$ \um\ agrees with our blackbody component explaining the IR bump, their interpretation was biased due to the lack of available FIR and sub-mm polarimetric observations. Thus, we here stress the need of FIR and sub-mm polarimetric observations to investigate the cores of AGN.

We compare our measured polarization and suggested interpretation with other AGN and astrophysical objects in the $10-100$ \um\ wavelength range. \citet{Siebenmorgen:2001aa} and \citet{2018MNRAS.tmp.1140L} found several highly polarized, $2-8$\%, AGN in the $7-14$ \um~wavelength range using the Infrared Space Observatory (ISO) and the GTC, respectively. At these wavelengths, AGN cores are known to be unpolarized, while high polarization arises from diffuse extended emission \citep{2018MNRAS.tmp.1140L}. Polarization up to $\sim12$ \% at $60-110$ \um\ using the Kuiper Airborne Observatory have been found around star forming regions (i.e. OMC1 and W3) and the Galactic center radio arc \citep[i.e.][]{Novak:1997aa,Dotson:2000aa}. Theoretical models \citep[e.g.][]{Hildebrand:2000aa,2002ApJS..142...53V,2002MNRAS.329..647A} show that the FIR polarization arises from emission of aligned dust grains with several magnetic field configurations. These results support that the observed high polarized Cygnus A core within the $10-100$ \um\ arises from aligned dust grain emission. Further FIR polarimetric and deeper observations of a larger AGN sample and wavelength coverage will provide statistically significant results to investigate how ordinary or extraordinary these findings are and to study the role of the magnetic field configuration.


\acknowledgments

We are grateful to Chris Packham and Edgar Ramirez for their numerous comments that helped to clarify and improve the text. Based on observations made with the NASA/DLR Stratospheric Observatory for Infrared Astronomy (SOFIA). SOFIA is jointly operated by the Universities Space Research Association, Inc. (USRA), under NASA contract NAS2-97001, and the Deutsches SOFIA Institut (DSI) under DLR contract 50 OK 0901 to the University of Stuttgart. Financial support for this work was provided by NASA through award \#05\_0071 issued by USRA. Herschel is an ESA space observatory with science instruments provided by European-led Principal Investigator consortia and with important participation from NASA. E.L.-R. acknowledges support from the Japanese Society for the Promotion of Science (JSPS) through award PE17783, the National Observatory of Japan (NAOJ) at Mitaka and the Thirty Meter Telescope (TMT) Office at NAOJ-Mitaka for providing a space to work and great collaborations during the short stay in Japan.


\vspace{5mm}
\facilities{SOFIA (HAWC+), \textit{Herschel} (PACS, SPIRE)}


\software{astropy \citep{2013A&A...558A..33A}  
          }



\appendix

\section{Supporting material}\label{app:sup}

Table \ref{tab:table1} shows the total and polarized measurements of the nucleus of Cygnus A used in Section \ref{sec:sed}. Figure \ref{fig:fig4} shows the \textit{Herschel} images of Cygnus A.

%%%%%%%%%%%%%
%%%% TABLE 1 %%%%
%%%%%%%%%%%%%%

\begin{deluxetable*}{ccccccl}[ht]
\tablecaption{Total and polarized nuclear measurements of Cygnus A.\label{tab:table1}}
\tablecolumns{7}
\tablenum{1}
\tablewidth{0pt}
\tablehead{\colhead{$\lambda$} &	\colhead{Aperture}	& \colhead{Flux Density}	 &\colhead{P} & \colhead{PA} & \colhead{Polarized Flux} & \colhead{Reference} \\ 
\colhead{(\um)}	&	\colhead{(\arcsec)} & \colhead{(mJy)} &\colhead{(\%)} & \colhead{($^{\circ}$)} & \colhead{(mJy)}
}
\startdata
2.0	&	0.375 (375 pc)	&	$0.005\pm0.001$	&	$10\pm1.5$	&	$201\pm3$	& $0.49\pm0.04$	&	\citet{Tadhunter:2000aa}	\\
8.7	&	0.38 (380 pc)	&	$29\pm6$		&	$11\pm3$		&	$27\pm8$		& $3.19\pm0.66$	&	\citet{Lopez-Rodriguez:2014aa}	\\
11.6	&	0.38 (380 pc)	&	$45\pm9$		&	$12\pm3$		&	$35\pm8$		& $5.4\pm2.08$	&	\citet{Lopez-Rodriguez:2014aa}	\\
18.0	&	2.0 (2 kpc)	&	$319\pm27$	&	\dots	&	\dots	&	\dots	& \cite{2002ApJ...566..675R} \\
53	&	5.0 (5 kpc)	&	$2218\pm222$	&	$11\pm3$		&	$43\pm8$		& $244\pm98$	&	This work	(HAWC+)\\
70	&	6 (6 kpc)	& $2520\pm30$	&\dots	&\dots	&\dots 	&	This work (\textit{Herschel})\\
89	&	8.0 (8 kpc)	&	$2088\pm195$	&	$9\pm2$		&	$39\pm7$		& $188\pm63$	&	This work	(HAWC+)		 	\\
100 & 	7 (7 kpc)	& $2160\pm10$	&\dots	&\dots	&\dots 	&	This work (\textit{Herschel})\\
160 &	12 (12 kpc)	& $1260\pm20$	&\dots	&\dots	&\dots 	&	This work (\textit{Herschel})\\
250 &	21 (21 kpc)	&  $560\pm40$	&\dots	&\dots	&\dots 	&	This work (\textit{Herschel})\\
350 &	29 (29 kpc)	&  $390\pm90$	&\dots	&\dots	&\dots	&	This work (\textit{Herschel})\\
500 & 	41 (41 kpc)	&  $360\pm110$	&\dots	&\dots	&\dots 	&	This work (\textit{Herschel})\\
1303	&	1.0 (1.0 kpc)		&	$480\pm48$ &\dots	&\dots	&\dots 	& \cite{2004ApJ...614..115W} \\
1400		&	12 (12 kpc) 	&	\dots	&	null		&	\dots 	& $<3.9$	&  \cite{Ritacco2017} \\
2000		&	18.2 (18.2 kpc) 	&	\dots	&	null		&	\dots 	& $<1.5$	&  \cite{Ritacco2017} \\
3446	&	1.0 (1.0 kpc)		&	$1050\pm105$ &\dots	&\dots	&\dots 	& \cite{2004ApJ...614..115W} \\
19986	&	1.0 (1.0 kpc)		&	$1350\pm135$ &\dots	&\dots	&\dots 	& \cite{2004ApJ...614..115W} \\
59958	&	1.0 (1.0 kpc)		&	$880\pm88$ &\dots	&\dots	&\dots 	& \cite{2004ApJ...614..115W} 
\enddata
%\tablenotetext{a}{Measured and modeled photometry as described in Section \ref{subsec:pho}.}
%\tablenotetext{b}{Fractional contribution of the several components used in the spectral decomposition described in Section \ref{subsec:seddes}. We estimate a 5\% uncertainty for the fractional contribution of each component.}
\end{deluxetable*}

%%%%%%%%%%%%%
%%%% FIGURE 3 %%%%
%%%%%%%%%%%%%%

%% The "ht!" tells LaTeX to put the figure "here" first, at the "top" next
%% and to override the normal way of calculating a float position
\begin{figure*}[ht!]
\includegraphics[angle=0,scale=0.48]{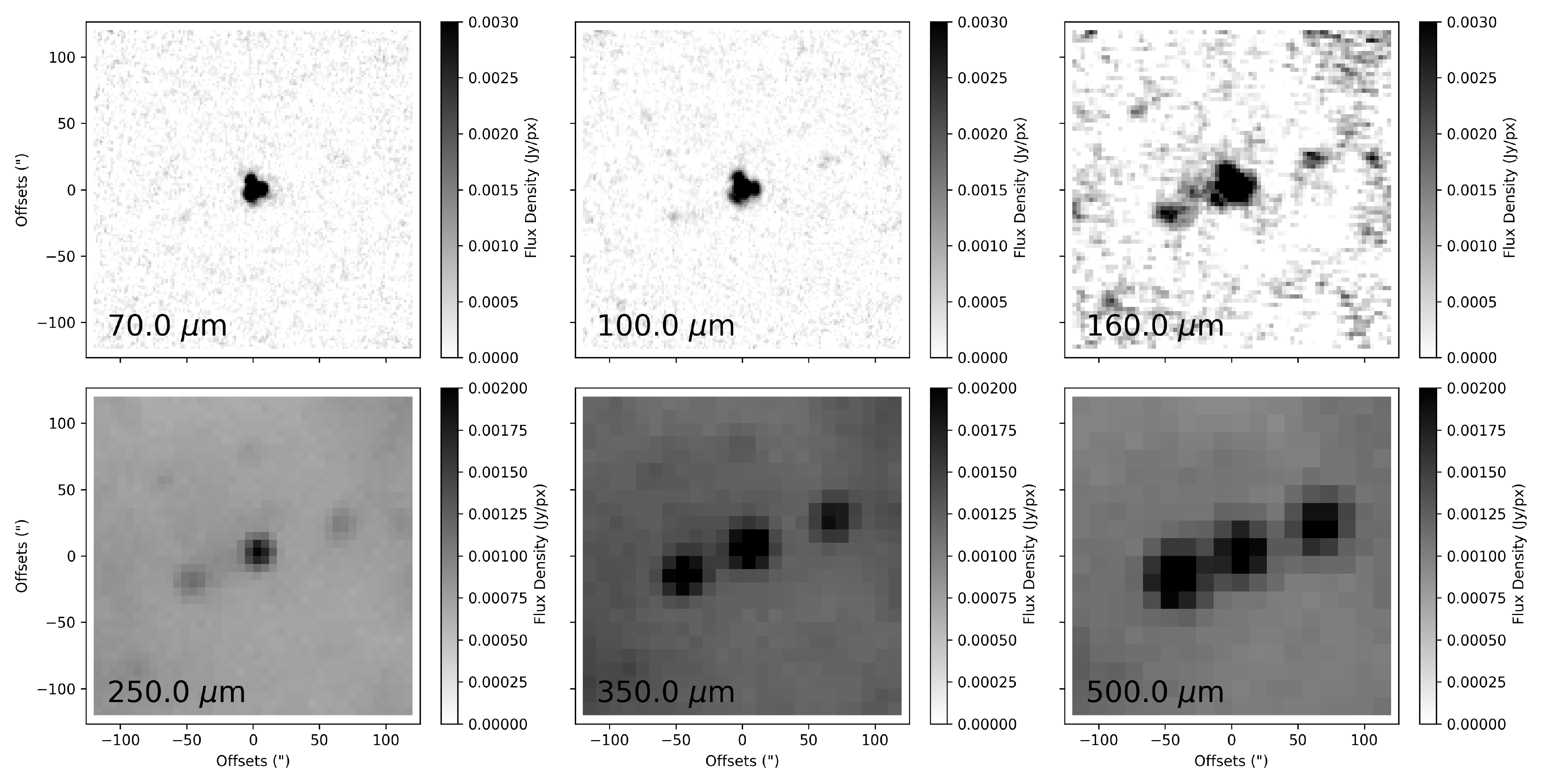}
\caption{\textit{Herschel} images of Cygnus A at 70, 100, 160, 250, 350 and 500 \um\ obtained with PACS and SPIRE. The field of view is $40 \times 40$ ($240 \times 240$ kpc$^{2}$)
\label{fig:fig3}}
\epsscale{2.}
\end{figure*}

%%%%%%%%%%%%%%%%%%%%%%%%%

\section{Polarization Model}\label{app:model}

Normalized Stokes qu are linear combinations of the dichroic absorption, emission and dust scattering profiles, such as:

\begin{eqnarray}
q(\lambda) = Af_{abs}(\lambda) + Bf_{em}(\lambda) + Cf_{sca}(\lambda) \\
u(\lambda) = Df_{abs}(\lambda) + Ef_{em}(\lambda) + Ff_{sca}(\lambda) \nonumber
\end{eqnarray}

\noindent 
where $A$, $B$, $C$, $D$, $E$, $F$ are scaling factors of the normalized absorptive, $f_{abs}(\lambda)$, emissive, $f_{abs}(\lambda)$ and dust scattering, $f_{sca}(\lambda)$, polarization components, that need to be estimated through a fitting routine. We used the fitting routine described by \citet{Lopez-Rodriguez:2016ab}, and set the scaling factors as uniform priors within the range of -0.3 to 0.3 of the normalized Stokes qu. Degree and P.A. of polarization are estimated as $P = \sqrt{q(\lambda)^2 + u(\lambda)^2}$, and $P.A. = 0.5 \times \arctan(u(\lambda)/q(\lambda))$ , respectively.

%%%%%%%%%%%%%
%%%% FIGURE 4 %%%%
%%%%%%%%%%%%%%

%% The "ht!" tells LaTeX to put the figure "here" first, at the "top" next
%% and to override the normal way of calculating a float position
\begin{figure*}[ht!]
\includegraphics[angle=0,scale=0.32]{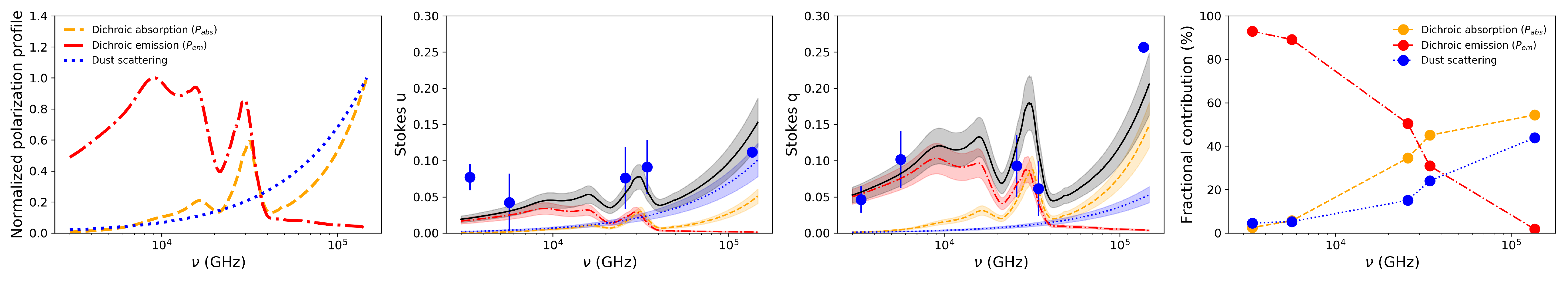}
\caption{Normalized polarization profiles of the absorptive (orange dashed line), emissive (red dotted-dashed line) and dust scattering (blue dotted line) components used to fit the observed Stokes qu of the nuclear polarization of Cygnus A in Fig. \ref{fig:fig2}. Fractional contribution (fourth panel) of the polarization mechanisms on our observations.
\label{fig:fig4}}
\epsscale{2.}
\end{figure*}

%%%%%%%%%%%%%%%%%%%%%%%%%

Polarization profiles were taken\footnote{The extinction curves can be found at \url{https://www.astro.princeton.edu/\~draine/dust/dustmix.html}} from the carbonaceous - silicate model for interstellar dust \citep{2001ApJ...548..296W}. The typical Galactic extinction of $R_{V} = 5.5$ was used. Given the uncertainties of our observations, we find little difference between the several values of $R_{V}$ in our final model. The absorptive polarization, $P_{abs}$, is the combination of a Serkowski curve \citep{Serkowski:1975aa} up to 5 \um, then the extinction curve, $\tau_{\lambda}$, is used. The emissive profile is estimated as $P_{em} = P_{abs}/\tau_{\lambda}$. For the dust scattering profile, the wavelength-dependent cross scattering sections for this model were used.  The best values of the scaling factors are $A = 0.053\pm0.011$, $B = 0.047\pm0.010$, $C = 0.187\pm0.041$, $D = 0.103\pm0.022$, $E = 0.095\pm0.021$, and $F = 0.103\pm0.022$.




\bibliography{CygA}

%\begin{thebibliography}{}
%\bibitem[Astropy Collaboration et al.(2013)]{2013A&A...558A..33A} Astropy Collaboration, Robitaille, T.~P., Tollerud, E.~J., et al.\ 2013, \aap, 558, A33 
%\end{thebibliography}



\end{document}